\documentclass[twocolumn, aps, prl, english]{revtex4-1}
\usepackage[T1]{fontenc}
\usepackage[latin9]{inputenc}
\usepackage{amsmath}
\usepackage{graphicx}
\usepackage{esint}

\makeatletter
\usepackage{babel}

\makeatother

\usepackage{color}   
\usepackage{hyperref}
\hypersetup{
    colorlinks=true, 
    linktoc=all,     
    linkcolor=blue,  
}

\begin{document}

\title{3D Quantum Anomalous Hall Effect in Hyperhoneycomb Lattices }

\author{Sang Wook Kim, Kangjun Seo, and Bruno Uchoa}

\affiliation{Department of Physics and astronomy, University of Oklahoma, Norman,
Oklahoma 73019, USA }

\date{\today}
\begin{abstract}
We address the role of short range interactions for spinless fermions
in the hyperhoneycomb lattice, a three dimensional (3D) structure
where all sites have a planar trigonal connectivity. For weak interactions,
the system is a line-node semimetal. In the presence of strong interactions,
we show that the system can be unstable to a 3D quantum anomalous
Hall phase with loop currents that break time reversal symmetry, as
in the Haldane model. We find that the low energy excitations of this
state are Weyl fermions connected by surface Fermi arcs. We show that
the 3D anomalous Hall conductivity is $e^{2}/(\sqrt{3}ah)$, with
$a$ the lattice constant. 
\end{abstract}
\maketitle
\emph{Introduction. $-$} The quantum Hall conductivity describes
dissipationless transport of electrons in a system that breaks time
reversal symmetry (TRS) due to an external applied magnetic field.
In two dimensions (2D), the current is carried through the edges \cite{Thouless1982},
and the Hall conductivity $\sigma_{xy}$ is quantized in units of
$e^{2}/h$. In three dimensions (3D), the Hall conductivity is not
universal and has an extra unit of inverse length. As shown by Halperin
\cite{Halperin1987}, the 3D conductivity tensor on a lattice has
the form $\sigma_{ij}=e^{2}/(2\pi h)\epsilon_{ijk}G_{k}$, where $G$
is a reciprocal lattice vector (it could be zero). The realization
of the 3D quantum Hall effect has been proposed in systems with very
anisotropic Fermi surfaces \cite{Balicas,McKerman,Bernevig}, or else
in line-node semimetals \cite{Guinea,Mullen2015,Moessner,Kim}, where
the Fermi surface has the form of a line of Dirac nodes \cite{Burkov,Lu,Yang,Rappe,Weng,Yu,Heykikila,Chen,Ezawa,Wang,Bian,Bian2,Chan,Xie,Li}.

Equally interesting would be to realize the 3D quantum anomalous Hall
(QAH) effect \cite{Xu,Xu2}, where the anomalous Hall conductivity
emerges from the topology of the 3D band structure in the absence
of Landau levels. The first proposal for a Chern insulator system
was the Haldane model \cite{Haldane1988} on the honeycomb lattice,
where loop currents break TRS and can produce a non-zero Chern number
in the bulk states. Hyperhoneycomb lattices have the same planar trigonal
connectivity of the honeycomb lattice (see Fig. 1a), and hence could
provide a natural system for the emergence of a 3D QAH conductivity.
This lattice has been experimentally realized in honeycomb iridates
\cite{Modic} as a candidate for the Kitaev model \cite{Kitaev} and
awaits to be realized as a line-node semimetal.

In this Letter, we describe the 3D QAH state that emerges from interactions
in a hyperhoneycomb lattice with spinless fermions. This state competes
with a CDW state, and produces a very anisotropic gap around a line
of Dirac nodes in the semimetallic state. Due to a broken inversion
symmetry, the QAH gap changes sign along the nodal line, forming Weyl
points connected by Fermi arcs \cite{Wan,Armitage}. We show that
the QAH conductivity of the surface states is $e^{2}/(\sqrt{3}ah)$,
with $a$ the lattice constant.

\begin{figure}[b]
\vspace{-0.55cm}
\includegraphics[width=0.84\linewidth]{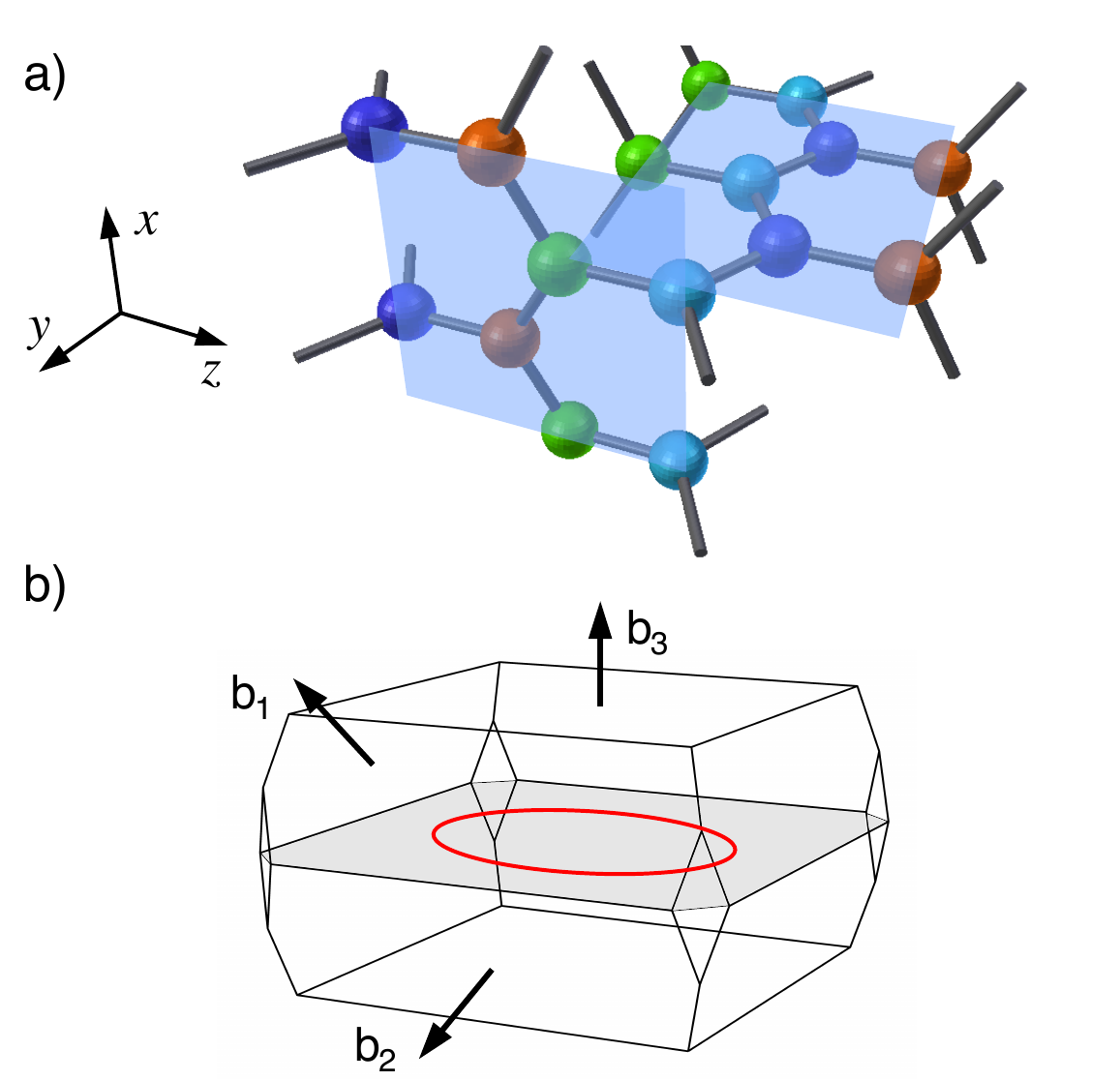}

\caption{(color online) a) Hyperhoneycomb lattice with four sublattices, indicated
by the different color sites. All sites are trigonally connected with
planar links spaced by $120^{\circ}$. The two planes are rotated
by $\pi/2$ along the $z$ direction, which has a screw symmetry.
b) 3D Brillouin zone of the crystal. In the semi-metallic state, a
closed zero energy line of Dirac nodes (Dirac loop) is shown in the
red curve on the $k_{z}=0$ plane (gray area). The black arrows indicate
the reciprocal lattice vectors. \label{fig:unit-cell}}
\end{figure}

\emph{Lattice model.$-$} We start from\emph{ }the tight binding model
of the hyperhoneycomb lattice, which has four atoms per unit cell
and planar links spaced by $120^{\circ}$, as shown in Fig. 1a. The
lattice has three vector generators $\mathbf{a}_{1}=(\sqrt{3},0,0)$,
$\mathbf{a}_{2}=(0,\sqrt{3},0)$ and $\mathbf{a}_{3}=(-\sqrt{3}/2,\sqrt{3}/2,3)$,
and the corresponding reciprocal lattice vectors $\mathbf{b}_{1}=(2\pi/\sqrt{3},0,-\pi/3)$,
$\mathbf{b}_{2}=(0,-2\pi/\sqrt{3},\pi/3)$ and $\mathbf{b}_{3}=(0,0,2\pi/3)$.
For a model of spinless fermions, which could physically result from
a strong Rashba spin orbit coupling, the kinetic energy is $\mathcal{H}_{0}=-t\sum_{\left\langle i,j\right\rangle }(a_{i}^{\dagger}a_{j}+h.c.)$,
where $a_{i}$ destroys an electron on site $i$, $t$ is the hopping
energy and $\langle ij\rangle$ denotes nearest neighbor (NN) sites.
In the four-sublattice basis, the Hamiltonian is a 4$\times$4 matrix
\cite{Mullen2015} 
\begin{equation}
\mathcal{H}_{0}=-t\left(\begin{array}{cccc}
0 & \Theta_{x} & 0 & \text{e}^{ik_{z}}\\
\Theta_{x}^{*} & 0 & \text{e}^{-ik_{z}} & 0\\
0 & \text{e}^{ik_{z}} & 0 & \Theta_{y}\\
\text{e}^{-ik_{z}} & 0 & \Theta_{y}^{*} & 0
\end{array}\right),\label{eq:Ho}
\end{equation}
where $\Theta_{\gamma}\equiv2\text{e}^{ik_{z}/2}\cos(\sqrt{3}k_{\gamma}/2)$,
with $\gamma=x,\,y$, and \textbf{$\mathbf{k}$} is the momentum away
from the center of the Brillouin zone (BZ). The electronic structure
has a doubly degenerate zero energy line of nodes in the form of a
Dirac loop at the $k_{z}=0$ plane, \textbf{$\mathbf{k}_{0}(s)\equiv(k_{x}(s),k_{y}(s),0)$}
in some parametrization that satisfies the equation $4\cos(\sqrt{3}k_{x}(s)/2)\cos(\sqrt{3}k_{y}(s)/2)=1$,
as schematically depicted in Fig. 1b. The projected low energy Hamiltonian
has the form 
\begin{equation}
\mathcal{H}_{0,p}(\mathbf{q})=\left[v_{x}(s)q_{x}+v_{y}(s)q_{y}\right]\sigma_{x}+v_{z}(s)q_{z}\sigma_{y},\label{eq:2}
\end{equation}
where $\mathbf{q}\equiv\mathbf{k}-\mathbf{k}_{0}(s)$ is the momentum
away from the nodal line, $\sigma_{x},\,\sigma_{y}$ are Pauli matrices,
with $v_{x}(s)=\frac{\sqrt{3}}{2}t\sin(\sqrt{3}k_{x}(s)/2)/(1+\alpha^{2})$,
$v_{y}(s)=\frac{\sqrt{3}}{2}\alpha^{2}t\sin(\sqrt{3}k_{y}(s)/2)/(1+\alpha^{2})$
and $v_{z}=-3t\alpha/(1+\alpha^{2})$ the quasiparticle velocities,
and $\alpha(s)=2\cos(\sqrt{3}k_{x}(s)/2)$. Hamiltonian (\ref{eq:2})
corresponds to the low energy spectrum 
\begin{equation}
\epsilon_{0}(\mathbf{q})=\sqrt{(v_{x}q_{x}+v_{y}q_{y})^{2}+v_{z}^{2}q_{z}^{2}},\label{epsilon0}
\end{equation}
that is gapless along the nodal line.

\begin{figure}
\includegraphics[width=0.95\linewidth]{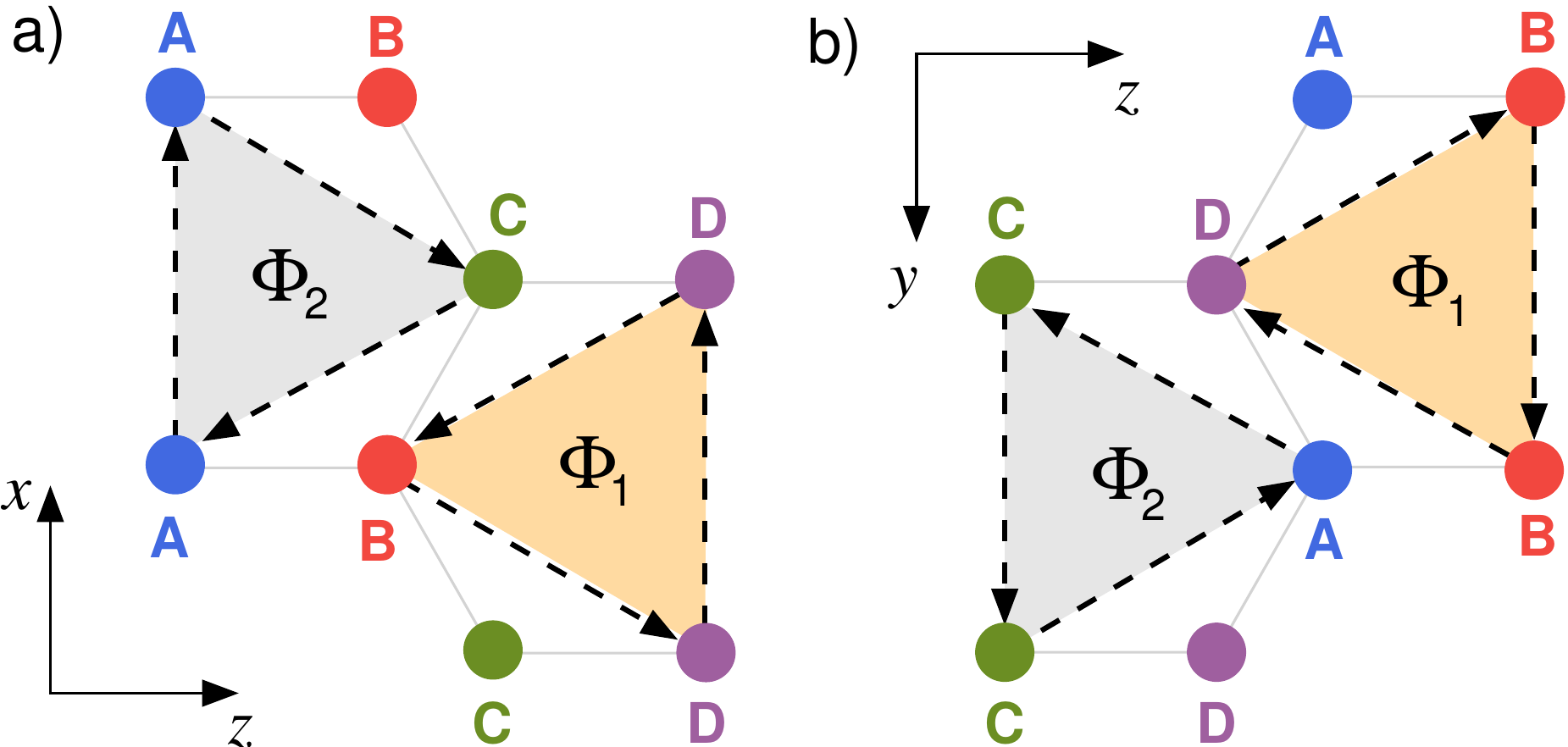}\caption{(color online) $xz$ (a) and $yz$ (b) planes of the hyperhoneycomb
lattice, with sublattices $A,\,B\,,C$ and $D$. Complex NNN hopping
terms $\chi_{ij}$ give rise to current loops with flux $\Phi$. The
lowest energy state has $\Phi_{1}=-\Phi_{2}$, which corresponds to
a zero total flux in the unit cell, with purely imaginary $\chi_{ij}$.
\label{fig:Phase-diagram-1}}
\end{figure}

The total Hamiltonian is $\mathcal{H}=\mathcal{H}_{0}+\mathcal{H}_{I}$,
where 
\begin{equation}
\mathcal{H}_{I}=V_{1}\!\sum_{\left\langle i,j\right\rangle }(\hat{n}_{i}-1)(\hat{n}_{j}-1)+V_{2}\negthickspace\!\sum_{\left\langle \left\langle i,j\right\rangle \right\rangle }(\hat{n}_{i}-1)(\hat{n}_{j}-1),\label{HI}
\end{equation}
is the interaction term, with $\hat{n}_{i}=a_{i}^{\dagger}a_{i}$
the density operator on site $i$, and $V_{1}$ and $V_{2}$ are the
repulsion between NN and next-nearest neighbors (NNN) sites, respectively.
For spinless fermions, one possible instability is a charge density
wave (CDW) state that corresponds to a charge imbalance among the
different sublattices. The CDW state is defined by the four component
order parameter $\rho_{\alpha}=\langle a_{i}^{\dagger}a_{i}\rangle-\rho_{0}$
with $i\in\alpha$ belonging to sublattice $\alpha=A\,,B,\,C,\,D$,
as shown in Fig 2, and $\rho_{0}$ a uniform density. At the neutrality
point, the local densities at the four sites of the unit cell add
up to zero, $\sum_{\alpha}\rho_{\alpha}=0$. The nodal line is protected
by a combination of TRS and mirror symmetry along the $z$ axis. The
state where $\rho_{A}=-\rho_{B}=\rho_{C}=-\rho_{D}$, namely $(\rho,-\rho,\rho,-\rho)$,
breaks the mirror symmetry and opens the largest gap among all possible
charge neutral configurations of $\rho_{\alpha}$. The more symmetric
state $(\rho,\rho,-\rho,-\rho)$ does not open a gap. Hence, the former
state is the dominant CDW instability. We will not consider other
possible states that enlarge the size of the unit cell \cite{Grushin},
such as an $n$-site CDW state, with $n>4$.

The other dominant instability is the QAH state, where complex hopping
terms between NNN sites lead to loop currents in the $xz$ and $yz$
planes, as shown in Fig. 2. Each plane can have loop currents with
opposite flux ($\Phi$), producing zero magnetic flux in the unit
cell, in analogy with the 2D case in the honeycomb lattice \cite{Haldane1988}.
The QAH order parameter is defined as $\chi_{ij}=\langle a_{i}^{\dagger}a_{j}\rangle$,
where $i$ and $j$ sites are connected by NNN vectors \cite{Raghu2008}.
We define the Ansatz $\chi_{ij}=\chi\text{e}^{i\phi_{ij}}$ for $i,\,j\in\{A,\,C\}$
sublattices and $\chi_{ij}=\chi\text{e}^{i\bar{\phi}_{ij}}$ for $i,\,j\in\{B,D\}$,
where $\chi$ is real. Due to particle-hole symmetry, $\chi_{ij}$
is purely imaginary and hence $\phi,\bar{\phi}=\pm\frac{\pi}{2}$.
The state that minimizes the free energy of the system has total zero
flux in the unit cell, $\Phi_{1}=-\Phi_{2}$ (see Fig. 2), when the
magnetic flux lines can more easily close. The QAH order parameter
is $\chi_{ij}=\pm i\chi$ for NNN sites and zero otherwise, with the
$+$ sign following the convention of the arrows in Fig. 2.

We perform a mean-field decomposition of the NN interaction in the
CDW state ($\rho$) and of the NNN repulsion in the QAH order parameter
$\chi_{ij}$. For simplicity, we absorb the couplings $V_{1}$ and
$V_{2}$ in the definition of the order parameters, $\rho V_{1}\to\rho$
and $\chi V_{2}\to\chi$, which have units of energy from now on.
The effective interaction in\emph{ }the four-sublattice basis is 
\begin{equation}
\mathcal{H}_{I}^{{\rm MF}}=\left(\begin{array}{cccc}
\chi g-3\rho & 0 & -\chi f & 0\\
0 & -\chi g+3\rho & 0 & \chi f^{*}\\
-\chi f^{*} & 0 & \chi g-3\rho & 0\\
0 & \chi f & 0 & -g\chi+3\rho
\end{array}\right)\!,\label{eq:Ho-1}
\end{equation}
where 
\begin{equation}
g(\mathbf{k})=2\left[\sin\left(\sqrt{3}k_{x}\right)+\sin\left(\sqrt{3}k_{y}\right)\right],\label{g}
\end{equation}
and 
\begin{equation}
f(\mathbf{k})=2\left[\text{e}^{i3k_{z}/2}\sin\left(\sqrt{3}k_{x}/2\right)+\text{e}^{-i3k_{z}/2}\sin\left(\sqrt{3}k_{y}/2\right)\right].\label{f}
\end{equation}
The mean-field Hamiltonian $\mathcal{H}^{\text{MF}}=\mathcal{H}_{0}+\mathcal{H}_{I}^{\text{MF}}$
has an additional constant energy term $E_{0}=6\rho^{2}/V_{1}+16\chi^{2}/V_{2}$
that is reminiscent of the decomposition of the interactions to quadratic
form. 

The phase diagram follows from the numerical minimization of the free
energy $F$ with respect to $\rho$ and $\chi$ at zero temperature,
$\partial F/\partial\chi=\partial F/\rho=0$. The semimetal state
is unstable to a CDW order at the critical coupling $V_{1,c}=0.41t$,
and to a QAH phase at $V_{2,c}=1.51t$. The CDW and QAH states compete
with each other, as shown in Fig. 3. Fluctuation effects are expected
to be less dramatic in 3D compared to the more conventional 2D case
\cite{Raghu2008,Motruck,Scherer}. Hence, the mean-field phase diagram
is likely a reliable indication of the true instabilities of the fermionic
lattice for the spinless case.

In real crystals, screening and elastic effects lead to a distortion
of the lattice in the CDW state, in order to minimize the Coulomb
energy due to electron-ion coupling, which can be high \cite{Cowley}.
While the CDW appears to be the leading instability over the QAH state,
the elastic energy cost to displace the ions and equilibrate the charge
in the electron-ion system may hinder the CDW order and favor the
QAH phase when $V_{2}>V_{2,c}$.

\begin{figure}[t]
\includegraphics[width=0.56\linewidth]{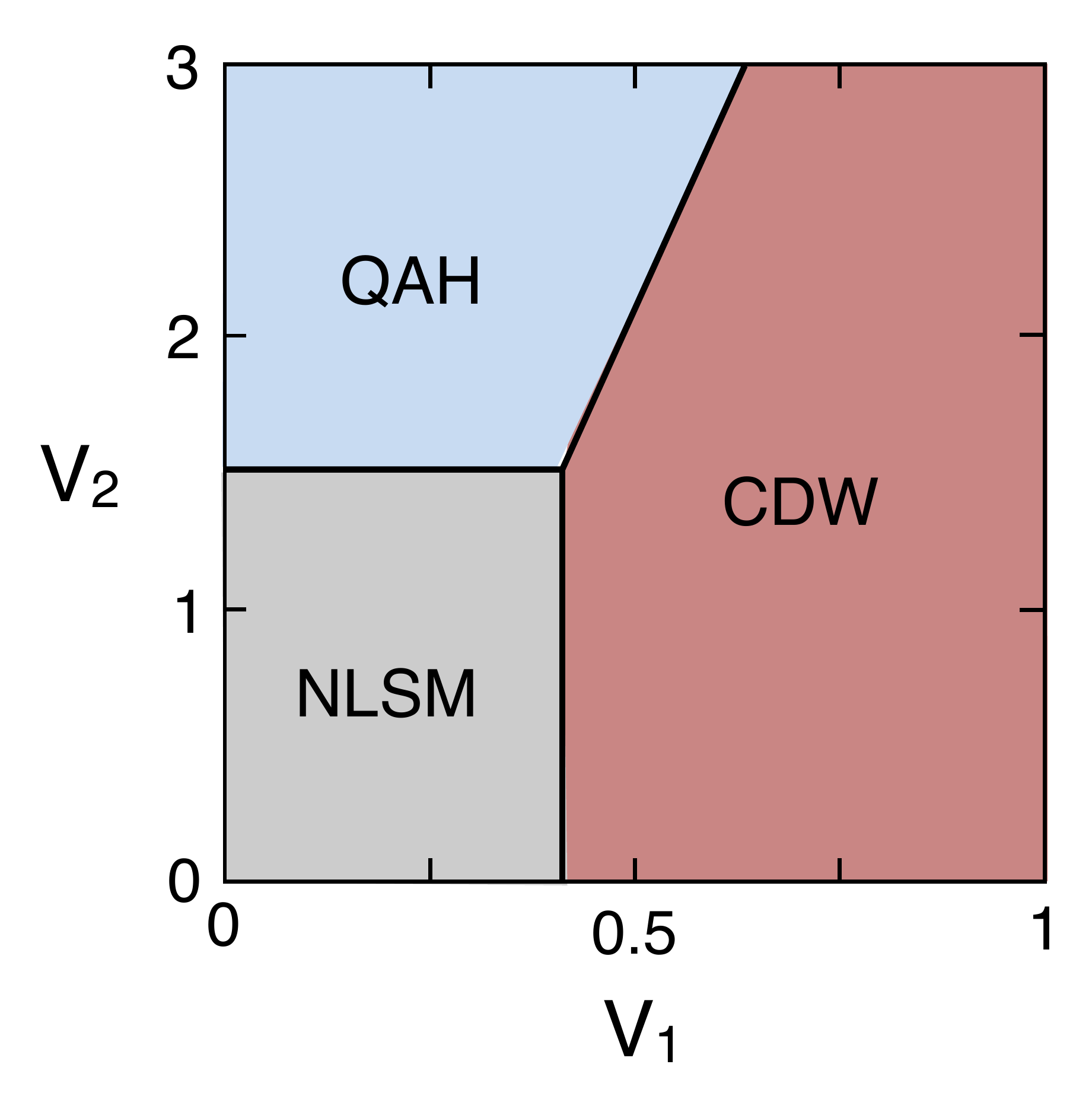}\vspace{-0.3cm}
\caption{(color online) a) Mean field phase diagram for spinless fermions.
The node-line semimetal phase (NLSM) turns into the CDW state at the
critical value $V_{1}=0.41t$ and into the QAH phase in 3D at $V_{2}=1.51t$.
The CDW is fully gapped, while the QAH phase has nodes around the
Dirac loop. \label{fig:Phase-diagram}}
\end{figure}

\emph{Low energy Hamiltonian}$.-$Integrating out the two high energy
bands using perturbation theory, the effective low energy Hamiltonian
(\ref{eq:2}) of the nodal line becomes massive, as expected. The
leading correction to Hamiltonian (2) around the nodal line to lowest
order in $\rho$ and $\chi$ has the form of a mass term 
\begin{equation}
\mathcal{H}_{I,p}\left(\mathbf{q}\right)=-\left[3\rho+m(\mathbf{k}_{0})+v_{x}^{\prime}q_{x}+v_{y}^{\prime}q_{y}\right]\sigma_{3},\label{eq:deltaH}
\end{equation}
where 
\begin{equation}
m(\mathbf{k}_{0})=\chi\left(g(\mathbf{k}_{0})+\frac{2}{\alpha+\frac{1}{\alpha}}f(\mathbf{k}_{0})\right)\label{eq:m}
\end{equation}
gives the QAH mass at the nodal line, with $v_{\gamma}^{\prime}(s)=2\chi\{\cos(\sqrt{3}k_{\gamma}(s))+\frac{1}{\alpha+1/\alpha}\cos(\sqrt{3}k_{\gamma}(s)/2)\}$
and $\alpha(s)$ defined below Eq. (\ref{eq:2}). The low energy spectrum
is 
\begin{equation}
\epsilon(\mathbf{q})=\pm\sqrt{\epsilon_{0}^{2}(\mathbf{q})+\left[3\rho+m(\mathbf{k}_{0})+v_{x}^{\prime}q_{x}+v_{y}^{\prime}q_{y}\right]^{2}},\label{eq:epsilon}
\end{equation}
which describes either a uniformly gapped state in the CDW phase ($\rho\neq0$,
$\chi=0$) or a non-uniform QAH gap $(\rho=0,\chi\neq0)$ with six
nodes at the zeros of $m(\mathbf{k}_{0})$, as indicated in Fig. 4.

The CDW state breaks mirror symmetry along the $z$ axis, but preserves
the screw axis symmetry and hence creates a fully gapped state that
is rotationally symmetric along the nodal line. The QAH state on the
order hand breaks inversion symmetry. The mass term (\ref{eq:m})
changes sign at six zeros along the nodal line, as shown in Fig. 4b.
Two zeros are located along the diagonal direction of the nodal line,
at the points $\pm\mathbf{Q}_{1}=\pm\left(-\frac{2\pi}{3\sqrt{3}},\frac{2\pi}{3\sqrt{3}},0\right)$.
The other four zeros of $m(\mathbf{k}_{0})$ are symmetrically located
around that direction, at $\pm\mathbf{Q}_{2}=\pm(Q_{+},Q_{-},0)$
and $\pm\mathbf{Q}_{3}=\mp(Q_{-},Q_{+},0)$, as shown in Fig. 4, with
$Q_{\pm}=\frac{1}{\sqrt{3}}\text{arccos}(\frac{\sqrt{17}-1}{4})\pm\frac{1}{\sqrt{3}}\arccos(\frac{3-\sqrt{17}}{4})$.
The position of the nodal points extracted from the low energy Hamiltonian
(\ref{eq:deltaH}) is in agreement with the values calculated numerically
from Hamiltonians (\ref{eq:Ho}) and (\ref{eq:Ho-1}) in the regime
where $\chi\ll t$. For larger values of $\chi$, the nodal points
$\pm\mathbf{Q}_{2}$ and $\pm\mathbf{Q}_{3}$ can move in the $k_{z}=0$
plane, as the position of the nodal line is renormalized by the interactions.
The two nodal points in the diagonal $\pm\mathbf{Q}_{1}$ remain fixed.

\begin{figure}[t]
\includegraphics[width=0.97\linewidth]{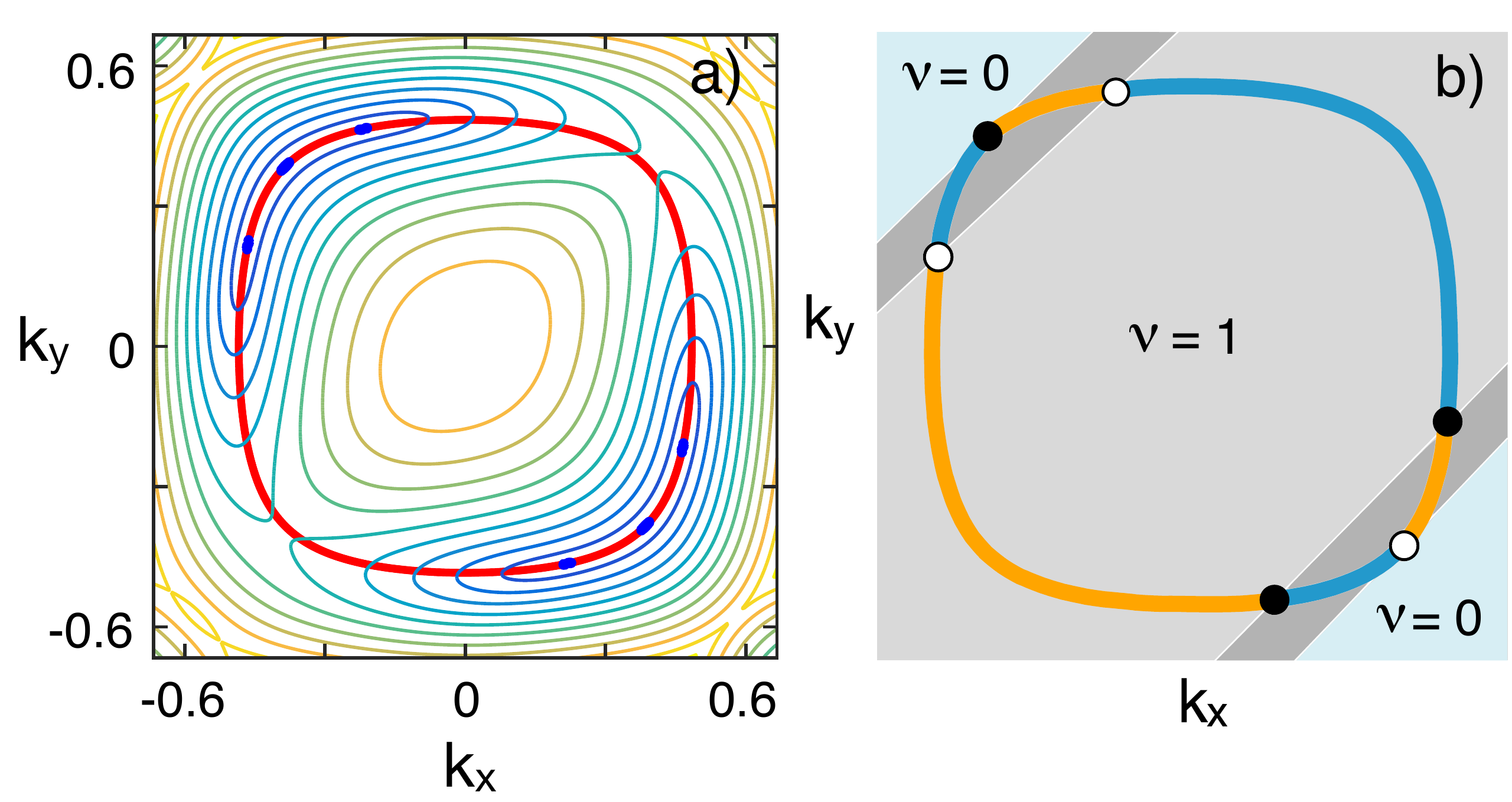}\vspace{-0.3cm}
\caption{(color online) a) Contour plot of the mass gap of the QAH state (\ref{eq:m})
around the Dirac loop (red line). Momenta are in units of $\pi$.
The gap vanishes at six points along the nodal line indicated by the
blue dots, where the contours collapse. b) Schematic picture of the
sign of the QAH gap around the nodal line: blue line ($m>0$); orange
line ($m<0$). At the nodes, the low energy excitations are Weyl fermions,
with helicities $\gamma=+1$ (black dots) and $\gamma=-1$ (white).
The Weyl points are intercepted by four planes oriented in the $(1\bar{1}0)$
direction (diagonal lines). Those planes form domain walls separating
slices of the BZ with distinct Chern numbers. Gray area: $\nu=1$.
Dark gray: $\nu=-1$. Light blue: $\nu=0$. \label{fig:mass gap}}
\end{figure}

\begin{figure*}
\includegraphics[width=0.81\linewidth]{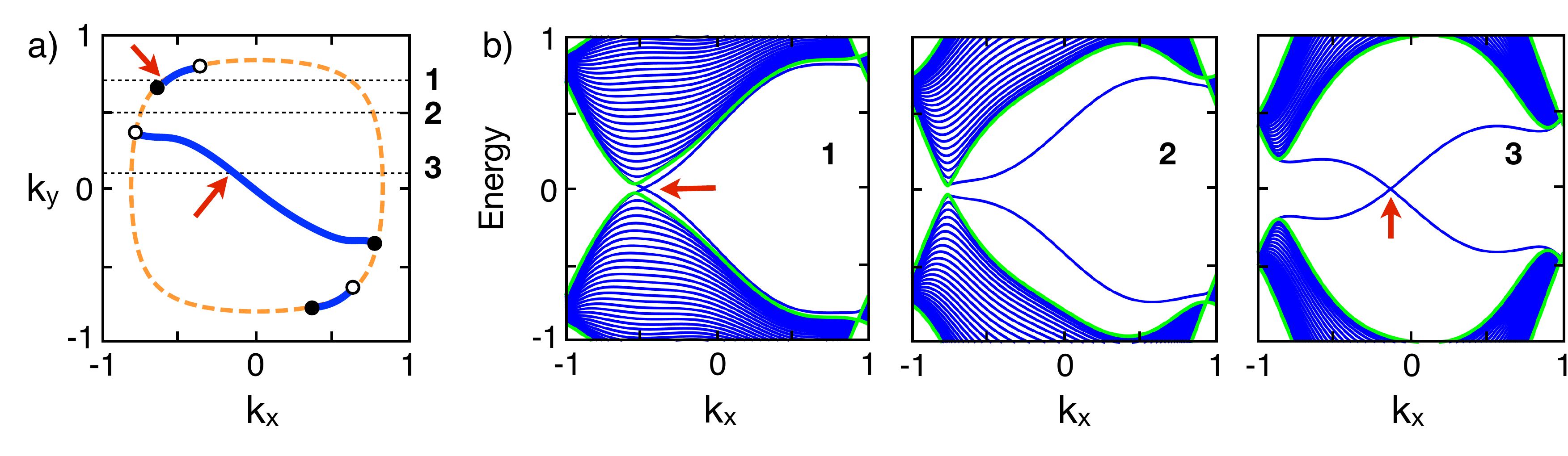}\caption{\label{fig:SChern-1}(color online) a) Fermi arcs on the (001) surface
BZ indicated by the blue solid lines. The brown dashed represents
the nodal ring in the bulk in the absence of interactions. The black
and white circles are the Weyl nodes in the bulk with positive and
negative helicities. b) Panels showing the energy dispersions $E(k_{x},k_{y})$
along the three momentum space cuts labeled $1,\,2,\,3$, indicated
in panel a. All momenta are in units of $\pi/\sqrt{3}$ and energy
has units of the hopping energy $t$. (left) cut 1, with $k_{y}=3\pi/(4\sqrt{3})$;
(center) cut $2$, with $k_{y}=\pi/(2\sqrt{3})$, where there is no
Fermi arc; (right) cut $3$, with $k_{y}=\pi/(10\sqrt{3})$. Red arrows
indicate the zero energy modes on the surface BZ. }
\end{figure*}

Expanding the the mass term around the zeros of $m(\mathbf{k}_{0})$,
the low energy quasiparticles around the nodes are Weyl fermions.
Performing a rotation of the quasiparticle momenta into a new basis
$p_{x}=(q_{x}-q_{y})/\sqrt{2},$ $p_{y}=-q_{z}$ and $p_{z}=(q_{x}+q_{y})/\sqrt{2}$,
the expansion around the the nodes at $\pm\mathbf{Q}_{1}$ gives the
low energy Hamiltonian 
\begin{equation}
\mathcal{H}_{\pm Q_{1}}(\mathbf{p})=\mathbf{h}_{\pm Q_{1}}(\mathbf{p})\cdot\vec{\sigma}=\sum_{i=x,y,z}v_{0,i}^{(\pm)}p_{i}\sigma_{i},\label{eq:hWeyl}
\end{equation}
with $\mathbf{p}$ the momentum away from the nodes and $v_{0,x}^{(\pm)}=\pm3\sqrt{2}t/4,\:v_{0,y}=3t/2$
and $v_{0,z}=\sqrt{3/2}\chi$ the corresponding velocities in the
rotated basis. The equation above describes two Weyl points with opposite
helicities $\gamma=(2\pi)^{-2}\int_{\Omega}\text{d}^{2}p\,\hat{h}\cdot(\partial_{p_{x}}\hat{h}\times\partial_{p_{y}}\hat{h})=\pm1$,
and hence broken TRS, with $\hat{h}=\mathbf{h}/|\mathbf{h}|$ a unitary
vector and $\Omega$ the surface of a small sphere enclosing each
node. Similarly, the expansion around the nodes $\pm\mathbf{Q}_{2}$
and $\pm\mathbf{Q}_{3}$ give Hamiltonians of Weyl fermions with helicities
$\pm1$, as indicated in Fig. 4b.

\emph{Anomalous Hall conductivity$.-$} The Weyl points delimit a
topological domain wall between slices of the BZ parallel to the $(1\bar{1}0)$
plane. Each slice in the light gray region in Fig. 4b crosses the
nodal line twice and has a well defined Chern number $\nu=+1$. The
slices in the dark gray regions across the domain walls have opposite
Chern number $\nu=-1$, as the QAH mass changes sign simultaneously
at the two Weyl points (with the same helicity) where each domain
wall intersects the nodal line. The BZ slices in the light blue region
do not cross the nodal line and have zero Chern number.

The 3D QAH conductivity is defined as $\sigma_{ij}=(e^{2}/h)(2\pi)^{-3}\int_{BZ}\text{d}^{3}k\sum_{n\in\textrm{filled}}(\frac{\partial}{\partial k_{i}}A_{j}-\frac{\partial}{\partial k_{j}}A_{i})$,
where $A_{j}=-i\left\langle \psi_{n}\right|\frac{\partial}{\partial k_{j}}\left|\psi_{n}\right\rangle $
is the Berry connection of the $n$-th occupied Block band integrated
over the entire BZ \cite{Haldane2}. For the hyperhoneycomb lattice
in the QAH state, 
\begin{equation}
\sigma_{ij}=\frac{e^{2}}{2\pi h}\int_{C}\text{d}k_{k}\epsilon_{ijk}\nu_{(k)}(\mathbf{k}_{0})=\frac{e^{2}}{2\pi h}\epsilon_{ijk}(\mathbf{b}_{1}+\mathbf{b}_{2})_{k},\label{eq:sigmaij}
\end{equation}
where $\mathbf{b}_{1}+\mathbf{b}_{2}=\left(2\pi/\sqrt{3},-2\pi/\sqrt{3},0\right)a^{-1}$
is a reciprocal lattice vector, restoring the lattice constant $a$.
$\nu_{(j)}(\mathbf{k}_{0})=0,\,\pm1$ is the Chern number of a slice
of the BZ oriented in the $j=x,y,z$ direction, intersecting the nodal
line $\mathbf{k}_{0}(s)$ at two points, and $C\in[k_{j,\text{min}}(s),k_{j,\text{max}}(s)]$.
Therefore, we find that 
\begin{equation}
\sigma_{yz}=\sigma_{xz}=e^{2}/(\sqrt{3}ha),\label{sigmayz}
\end{equation}
while $\sigma_{xy}=0$. In the 3D QAH phase, the bulk of the system
is a semimetal with topologically protected Weyl quasiparticles \cite{Xu},
while charge currents spontaneously emerge on the $[100]$ and $[010]$
surfaces of the crystal.

\emph{Surface states$.-$} The presence of Weyl points in the QAH
state implies in the existence of Fermi arcs on the surfaces of the
lattice, connecting nodes with opposite helicities. In Fig. 5a, we
numerically calculate the Fermi arcs in the (001) surface Brillouin
zone, as shown in the solid blue lines. The nodes at $\pm\mathbf{Q}_{2}$
are connected by a Fermi arc crossing the center of the BZ, while
the pair of nodes at $\mathbf{Q}_{1}$, $-\mathbf{Q}_{3}$ and $-\mathbf{Q}_{1}$,
$\mathbf{Q}_{3}$ are connected by short Fermi arcs directed along
the nodal line.

In Fig. 5b, we scan the energy spectrum of the $k_{z}=0$ plane along
the $k_{x}$ axis along three paths indicated by the dotted horizontal
lines in panel 5a. Line 1 ($k_{y}=3\pi/(4\sqrt{3}$) intersects a
Fermi arc close to the node at $\mathbf{Q}_{1}$, as indicated by
the arrow in the left panel of Fig. 5b, which has a zero energy crossing
in the vicinity of a node. The scan on line 2, at $k_{y}=\pi/(2\sqrt{3})$,
does not intercept a Fermi arc, as shown in the center panel of Fig.
5b. The third path at $k_{y}=\pi/(10\sqrt{3})$ crosses the Fermi
arc near the center of the zone, as indicated by the zero energy mode
shown in the right panel of Fig. 5b.

\emph{Conclusions$.-$ }We have shown that hyperhoneycomb lattices
with spinless fermions may host a 3D QAH effect, which competes with
a CDW state. The 3D anomalous Hall conductivity is $e^{2}/(\sqrt{3}ha)$.
Due to the symmetry of the mass term, which spontaneously breaks inversion
symmetry around the nodal line, the low energy excitations of the
QAH state have a rich structure, with Weyl fermions in bulk and topologically
protected surface states.

\emph{Acknowledgements.$-$ }We acknowledge F. Assaad, S. Parameswaran
and K. Mullen for helpful discussions. S. W. Kim and B.U. acknowledge
NSF CAREER Grant No. DMR- 1352604 for support. K. S. acknowledges
the University of Oklahoma for support.

\emph{Note.$-$ }Recently, we became aware of a related work \cite{Murakami},
which predicted the conditions for the emergence of Weyl points in
nodal-line semimetals from symmetry arguments.

\end{document}